\documentclass[12pt,a4paper]{article}
\usepackage{graphicx}
\usepackage{amsfonts}
\usepackage{amssymb}
\usepackage{dcolumn}
\usepackage{bm}
\usepackage{subeqnarray}
\newcommand{\beq}{\begin{equation}}
\newcommand{\eeq}{\end{equation}}
\newcommand{\bea}{\begin{eqnarray}}
\newcommand{\eea}{\end{eqnarray}}
\begin{document}

\begin{center}
{\bf \large Random phase approximation for the 1D anti-ferromagnetic Heisenberg model.}\\[2ex]

A.~Rabhi$^1$, P.~Schuck$^{2,3}$ and J.~Da~Provid\^encia$^1$ \\

$^1$ Departamento de F\'isica, Universidade de Coimbra, 3004-516 Coimbra, Portugal \\
$^2$ Institut de Physique Nucl\'eaire, 91406 Orsay Cedex, France  \\
$^3$ LPM2C, Maison des Magist\`eres-CNRS, 25, avenue des Martyrs, BP 166, 38042, Grenoble Cedex, France\\

\begin{abstract}
The Hartree-Fock-RPA approach is applied to the 1D anti-ferromagnetic Heisenberg model 
in the Jordan-Wigner representation. Somewhat contrary to expectation, this leads 
to reasonable results for spectral functions and sum rules in the symmetry unbroken phase.
In a preliminary application of Self-Consistent RPA to finite size chains strongly improved results are obtained.
\end{abstract}

pacs : 75.40.Gb, 75.50.Ee, 75.10.Pq, 75.10.Jm

\end{center}

The 1D Anti-Ferromagnetic Heisenberg Model (AFHM)~\cite{bet0} belongs to one of the most classic 
many body research fields. It was the first many-body model to be solved exactly by Bethe 
with his famous Bethe ansatz solution~\cite{bet1} and it has been a playground for testing 
and developing many body approaches ever since. In spite of tremendous progress in the 
understanding of many facets of the model, it continues to be a very active field of 
interest and research~\cite{mikes,boug}. 

In an attempt to apply extended RPA-like theories, as {\it e.g.} the recently 
developed Self-Consistent RPA (SCRPA) approach~\cite{scrp2, kru}, we found out, somewhat 
to our surprise, that even the standard RPA theory has so far not fully been developed
in its application to the 1D-AFHM. This probably stems from the fact that the HF-RPA 
scheme is usually considered as unreliable in low dimensions~\cite{taka}. 
However, it will be found that RPA shows interesting features even in this 1D-version 
of the model. On the other hand SCRPA shows very promising results in a number of 
cases~\cite{scrp2} and in a first application for a small numbers of sites we here get 
very close agreement with results from exact diagonalization.

We mostly consider the isotropic Heisenberg model with anti-ferromagnetic coupling, 
however some consideration will also be given to the anisotropic case, for instance 
to the XXZ-anisotropic Heisenberg Hamiltonian 
\beq
H=\sum^{N}_{n=1}\left\{\frac{1}{2}\left(S^{+}_{n+1}S^{-}_{n}+S^{-}_{n+1}S^{+}_{n}\right)
+gS^{z}_{n+1}S^{z}_{n}\right\}.
\eeq
where $S^{i}_{n}$ is the spin operator of the $n$-th site, $g$ is the anisotropy 
parameter and $N$ is the number of sites. Periodic boundary conditions ($S^{i}_{N+1} =S^{i}_{1}$) 
are applied. We will work with the AFH model in the Jordan-Wigner (JW) representation~\cite{JW}.
The Hamiltonian in momentum representation is given by~\cite{bet2}
\beq
H=\frac{gN}{4}+\sum_{k}\varepsilon^{o}_{k}\psi^{\dagger}_{k}\psi_{k}
+\frac{1}{4}\sum_{k_{1}k_{2}k_{3}k_{4}}\bar{v}_{k_{1}k_{2}k_{3}k_{4}}
\psi^{\dagger}_{k_{1}}\psi^{\dagger}_{k_{2}}\psi_{k_{4}}\psi_{k_{3}}
\label{ham}
\eeq
with
\beq
\varepsilon^{o}_{k}=\cos(k)-g
\eeq
and
\beq
\bar{v}_{k_{1}k_{2}k_{3}k_{4}}=\frac{2g}{N}\left(\cos(k_{1}-k_{3})
-\cos(k_{1}-k_{4})\right)\delta (k_{1}+k_{2}-k_{3}-k_{4}).
\eeq
The isotropic point is given by $g=1$ whereas for $g\neq 1$ we have the XXZ model. 
In order to take care of the boundary conditions the momentum sums run over the values 
$\displaystyle k=j\frac{\pi}{N}$ with :
\bea
j&=& \pm 1, \pm 3, \pm 5, \dots, \pm (N-1) \quad \mbox{for $n$ even} \cr
j&=& 0, \pm 2, \pm 4, \dots, \pm (N-2), N \quad \mbox{for $n$ odd}
\eea
where $n$ is the fermion number (in this note we only consider the sector $n=N/2$) and where we have assumed, 
without loss of generality, that $N$ is even. The Kronecker symbol is defined by :
\bea
\delta(p-p^{\prime}) &=& 1 \qquad \mbox{if} \quad p=p^{\prime}\pm 2 \pi \tau; \quad \tau=0,1,2,\cdots \cr
                     &=& 0 \qquad \mbox{otherwise}.
\eea

We first want to study the Hartree-Fock (HF) approximation, keeping translational invariance,
\textit{i.e.} momentum as a good quantum number. The JW-HF single particle energies are given by 
$
\epsilon_{k}=\langle\left\{\psi_{k},[H,\psi^{\dagger}_{k'}]\right\}\rangle.
$
Here $[..,..]$ stands for commutator, $\{..,..\}$ for anticommutator, and $\langle ... \rangle$ denotes
the expectation value in the HF state. One obtains
\beq
\epsilon_{k}=\varepsilon^{o}_{k}+\frac{2g}{N}\sum_{h}\left(1-\cos(k-h)\right).
\eeq
where the sum goes over the occupied states "h" only. The HF groundstate energy is then given by
\beq
E_{0}^{HF}=\frac{gN}{4}+\sum_{h}\left(\varepsilon^{o}_{h}+\frac{g}{N}\sum_{h'}\left(1-\cos(h-h')\right)\right).
\eeq
It is interesting to note that with respect to the N\'eel state the HF energy of the 
JW transformed isotropic AF Hamiltonian has actually a lower minimum. In table~\ref{tab1} 
we give the HF energy for the N\'eel state and the JW-HF energies
as a function of the number of sites $N$ for the isotropic situation with $g=1$.

\begin{table}[htb]
\begin{tabular}{c|cccc}
\hline
\hline
N & 4 & 6& 8 & 12  \\
\hline
N\'eel energy & -1.0  & -1.5 & -2.0 & -3.0 \\
\hline
JW-HF energy  & -1.9142  & -2.6667 & -3.4667 & -5.1077 \\
\hline
Deformed JW-HF energy & -1.9142  & -2.6667 & -3.4856 & -5.1927 \\
\hline
\end{tabular}
\caption{Mean-field energies with various approximations for some low number of site cases (see text for details, {\it e.g.} 'deformed JW-HF energy' stands for JW-HF energy in the symmetry broken phase).}
\label{tab1}
\end{table}

So the JW-HF theory takes into account more of the interaction energy than the N\'eel state in
real space. It seems, however, not easy to analyze the physical content in real space of the
JW-HF groundstate. 

\begin{figure}[htb]
\centering
\vspace{0.5 in}
\includegraphics[width=0.75\linewidth]{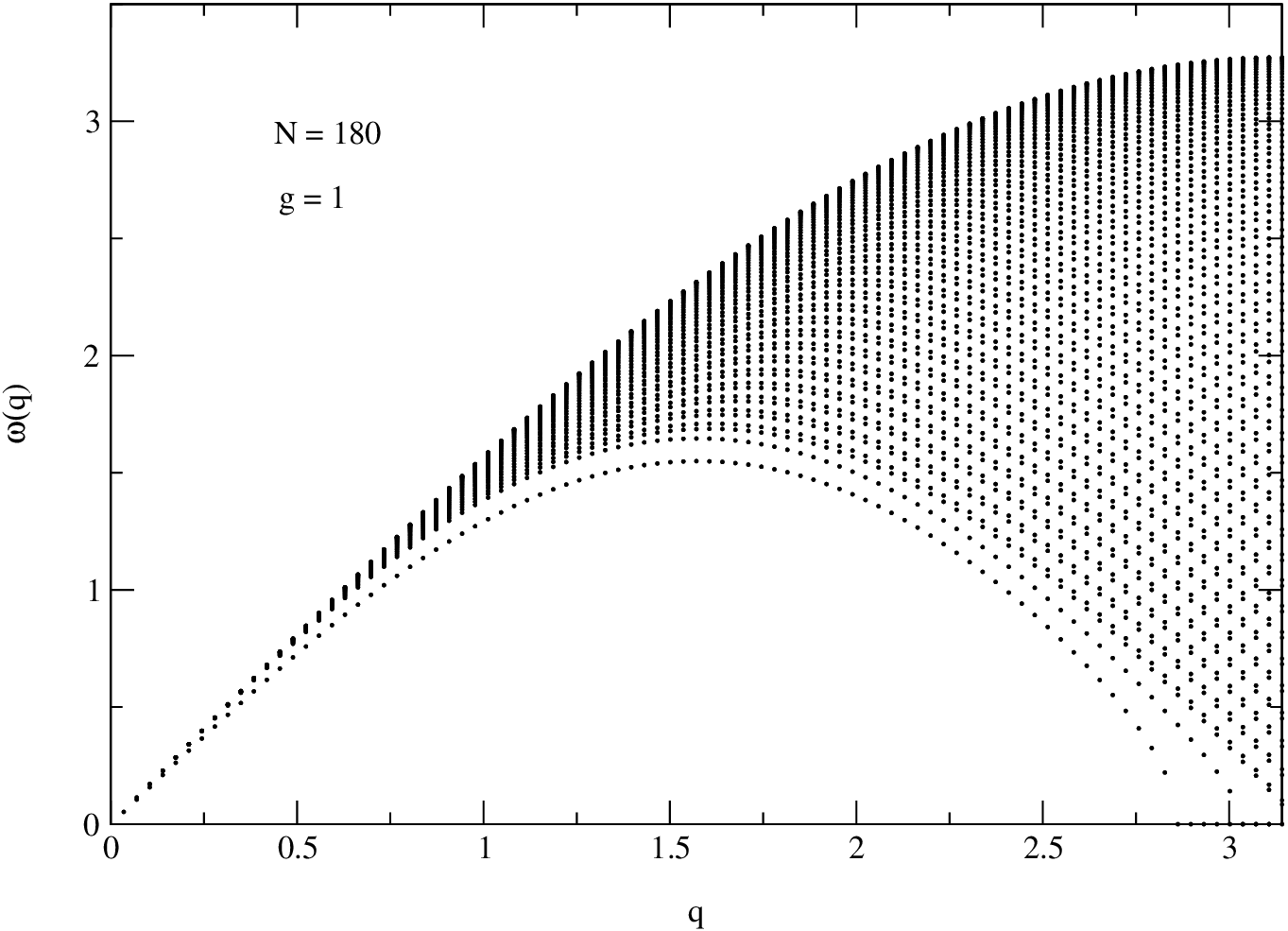}
\caption{RPA Excitation spectrum for $N=180$ and $g=1$.}
\label{fig-1}
\end{figure}

\begin{figure}[htb]
\centering
\vspace{0.5 in}
\includegraphics[width=0.75\linewidth]{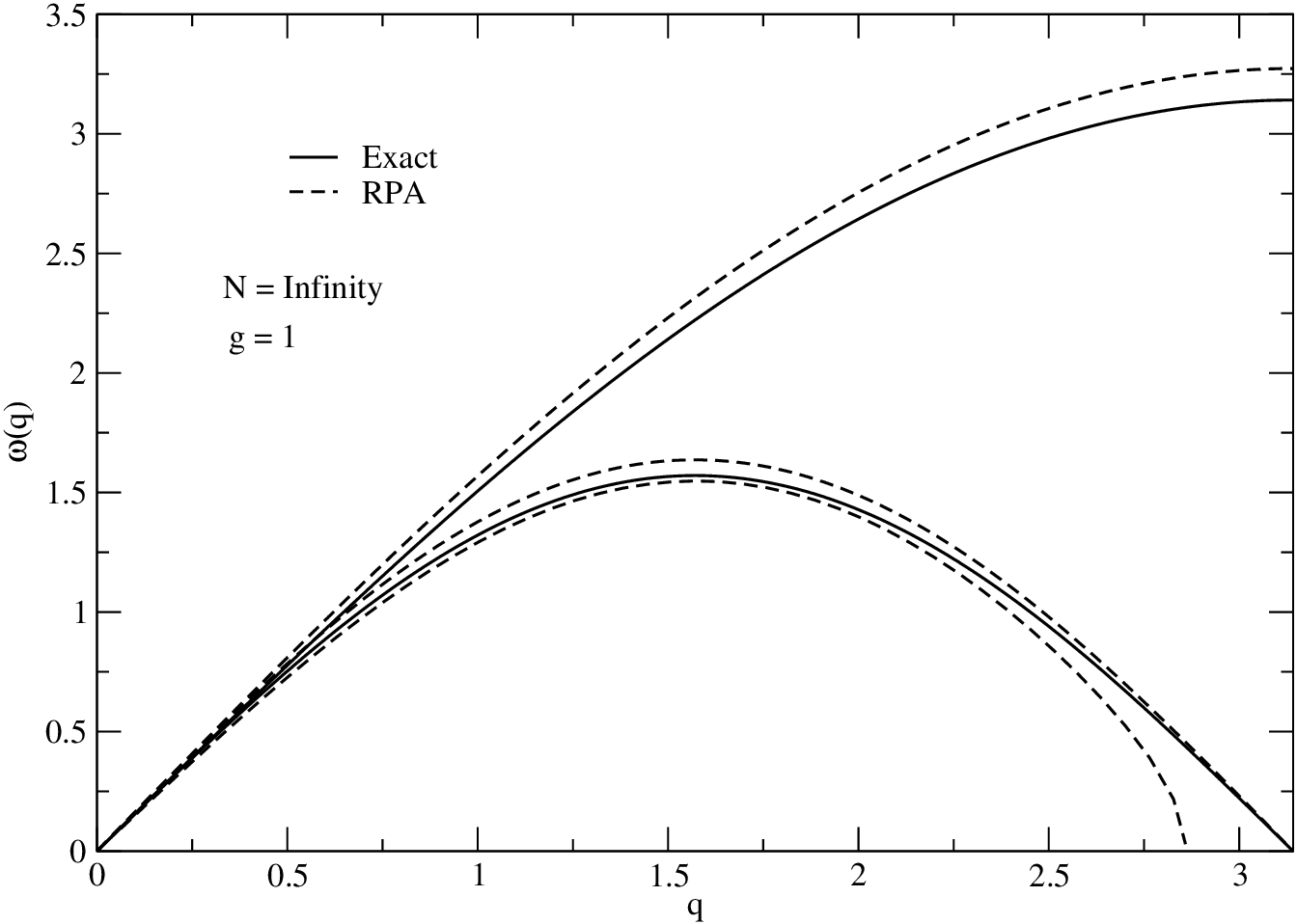}
\caption{Excitation spectrum for the infinite chain. The lowest broken line corresponds to the 
discrete RPA branch. The two upper broken lines limit the RPA p-h continuum. The two full lines 
limit the continuum in the exact two-spinon case~\cite{boug} where there is no discrete state.}
\label{fig-2}
\end{figure}

\begin{figure}[htb]
\centering
\vspace{0.5 in}
\includegraphics[width=0.75\linewidth]{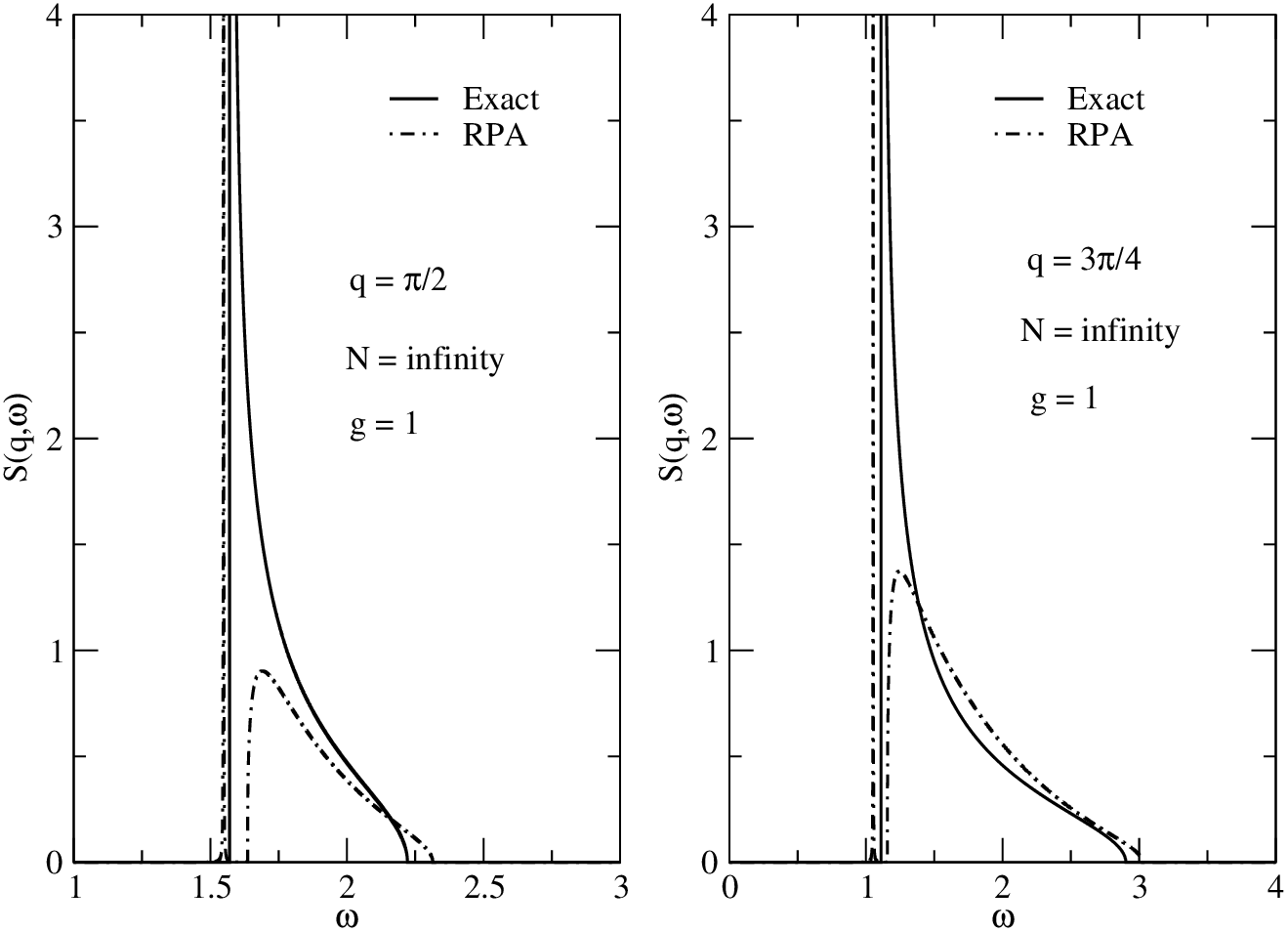}
\caption{Exact two-spinon and RPA dynamic spin structure factor, in the thermodynamic limit, for $\displaystyle q=\frac{\pi}{2}$ and $\displaystyle q=\frac{3\pi}{4}$.}
\label{fig-3}
\end{figure}

We now go one step further and investigate the Random Phase Approximation 
to describe excitations on top of the before considered JW-HF groundstate. 
For this we elaborate the RPA equations via the equation of motion (EOM)
method~\cite{row, RS}. We introduce an RPA particle(p)-hole(h) excitation operator
\beq
Q^{\dagger}_{\nu}=\sum_{ph}X^{\nu}_{ph}\psi^{\dagger}_{p}\psi_{h}
-Y^{\nu}_{ph}\psi^{\dagger}_{h}\psi_{p}
\label{q1}
\eeq
with $|\nu\rangle=Q^{\dagger}_{\nu}|0\rangle$ and $|\nu\rangle$, $|0\rangle$ 
the excited state and groundstate respectively. With the vacuum condition 
$Q_{\nu}|0\rangle=0$, one arrives as usual~\cite{tho} to the RPA eigenvalue equation
\beq
\sum_{p'h'}\pmatrix{A & B \cr
-B^{*} & -A^{*}}_{ph,p'h'}\pmatrix{X^{\nu}_{p'h'} \cr Y^{\nu}_{p'h'}}=
\Omega_{\nu}\pmatrix{X^{\nu}_{ph} \cr Y^{\nu}_{ph}}
\label{m1}
\eeq
with
\begin{subeqnarray}
A_{ph,p'h'}&=&\langle [\psi^{\dagger}_{h}\psi_{p},[H,\psi^{\dagger}_{p'}\psi_{h'}]]\rangle \cr
&=&(\epsilon_{p}-\epsilon_{h})\delta_{pp'}\delta_{hh'}
+\bar{v}_{ph',hp'} \\
B_{ph,p'h'}&=&-\langle [\psi^{\dagger}_{h}\psi_{p},[H,\psi^{\dagger}_{h'}\psi_{p'}]]\rangle
=\bar{v}_{pp',hh'} 
\label{m2}
\end{subeqnarray}
We get the standard RPA from Eqs.(\ref{m1}) in evaluating expectation values with the before 
elaborated JW-HF groundstate leading to the expressions given in Eq.(\ref{m2}). We show in Fig.~\ref{fig-1} 
for $N=180$ and $g=1$ the RPA eigenvalue spectrum as a function of the momentum transfer $|q|$.
We clearly see that there is a lower branch slightly detached from the more or less dense remainder 
{\it i.e.} the p-h continuum. Actually $N=180$ is already quite close to the thermodynamic limit 
which we also calculated using the Green function method following closely the work of Todani and Kawasaki~\cite{toda}. Here we only show the result in Fig.~\ref{fig-2}. 
We again see that a lower energy state (lowest broken line) becomes detached from the 
continuum, the latter being embraced by the two upper broken lines. The continuum was also studied by 
Todani and Kawasaki~\cite{toda} but for some reason they did not discuss the low-lying discrete state which 
is an important and prominent feature. In the thermodynamic limit the dispersion equation of 
the low-lying branch can be found from~(see Ref.~\cite{toda} Eq.(34)) 
\beq
1-f(q,\omega) \Re\left(G^{0}(q,\omega)\right)=0
\label{disper}
\eeq
where $G^{0}(q,\omega)$ is the renormalized particle-hole propagator~\cite{toda}  
\bea
G^{0}(q, \omega)=\frac{-1}{p\pi\sin(\frac{q}{2})}\frac{1}{2\sqrt{1-x^{2}}}
\log\left(\frac{\sqrt{1-x^{2}}+\sin(\frac{q}{2})}
{{\sqrt{1-x^{2}}}-\sin(\frac{q}{2})}
\right),
\eea
and
\beq
f(q,\omega) = 2\cos(q)-2px^{2}
\label{ff}
\eeq
with $\displaystyle x=\frac{\omega+i\eta}{2p\sin(\frac{q}{2})}$ and $\displaystyle p=1+\frac{2}{\pi}$.
From~Eq.~(\ref{disper}) also the lower $\omega_{L}$ and upper $\omega_{U}$ boundaries of 
the continuum can be obtained 
\begin{subeqnarray}
\omega_{L}=(1+\frac{2}{\pi})\sin(q) \\
\omega_{U}=2(1+\frac{2}{\pi})\sin(\frac{q}{2}).
\end{subeqnarray}
In the exact two-spinon case there is no discrete state and only a continuum limited by the upper 
$\displaystyle\left(\omega_{U}=\pi\sin(\frac{q}{2})\right)$ and lower $\displaystyle \left(\omega_{L}=\frac{\pi}{2}\sin(q)\right)$ full lines of Fig.~\ref{fig-2} exists~\cite{muller, boug}. 
The exact two-spinon dynamic structure factor diverges at the lower limit~\cite{muller, boug}. 
Nevertheless, globally, the RPA spectrum for the infinite chain resembles 
the exact solution, even though locally, especially at the lower edge, things go wrong even 
qualitatively. For instance, the fact that the excitation spectrum exhibits a low lying 
discrete state detached from the continuum is of course an artefact of the RPA approach. 
This fact has also very briefly been mentioned in Ref.~\cite{schneider} but no details 
are given there. We want to point out, however, and this is the main message of the present 
paper, that the RPA spectrum is already qualitatively quite similar to the exact one as seen in Fig.~\ref{fig-3} 
where we show the structure functions for $\displaystyle q=\frac{\pi}{2}$ and $\displaystyle q=\frac{3\pi}{4}$ 
corresponding to RPA and exact two-spinon calculation~\cite{boug}. The detachment of the discrete state is in fact only very mild and we therefore conclude that RPA may be a good starting point for more elaborate theories, even for situations where, a priori, it should not work well as for instance in 1D situation, as considered here.

Let us also mention that in the attractive (ferromagnetic) case quite naturally a discrete state, detached from the continuum, exists, even in the exact solution~\cite{beck-mull, schneider}.

\begin{figure}[htb]
\centering
\vspace{0.5 in}
\includegraphics[width=0.75\linewidth]{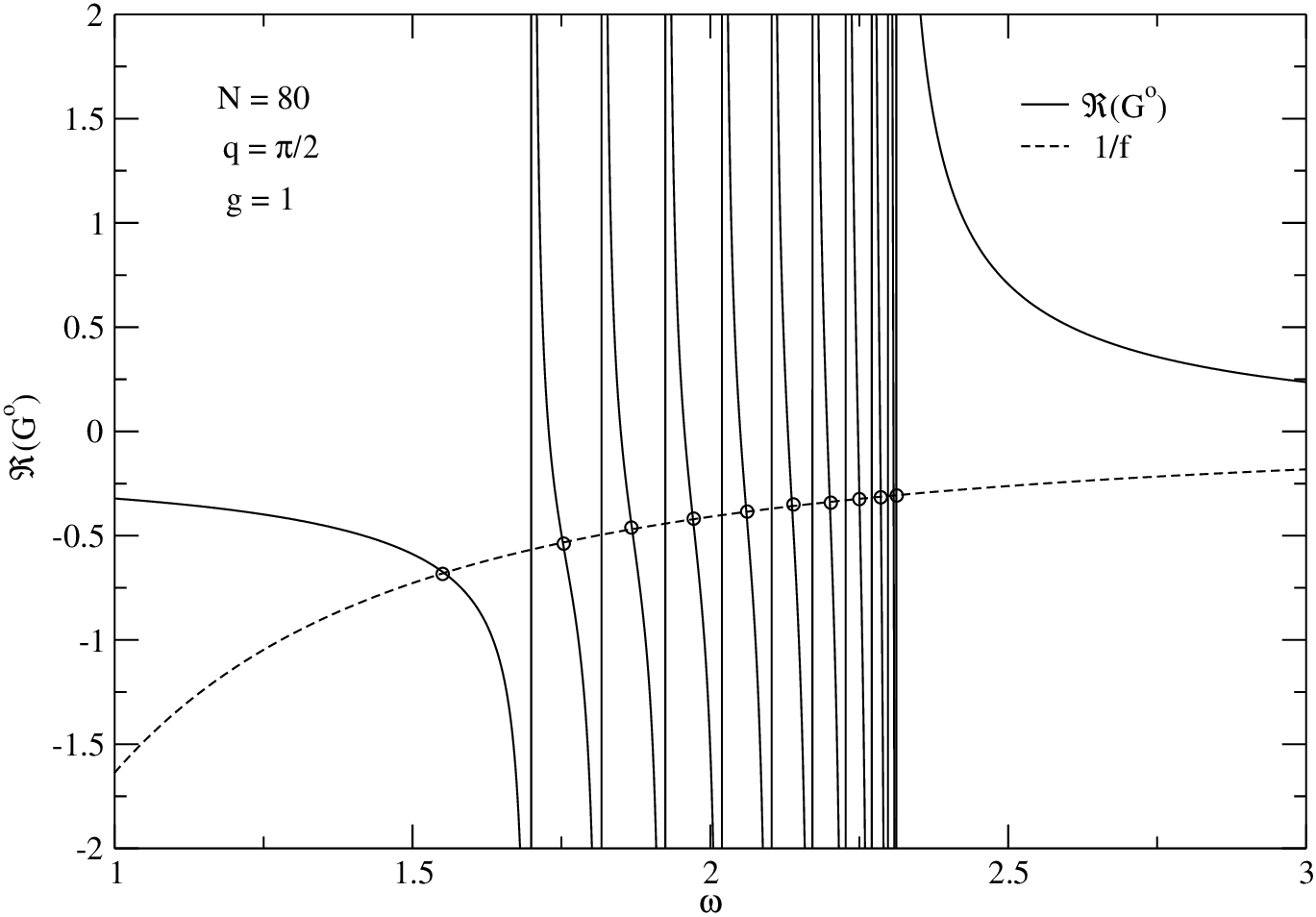}
\caption{Graphical solution of the dispersion relation for $N=80$ and $\displaystyle q=\frac{\pi}{2}$.}
\label{fig-4}
\end{figure}

To understand the reason why there exists in RPA a discrete collective state detached from the continuum it is instructive to solve Eq.~(\ref{disper}) for the dispersion relations $\omega(q)$ in the case of a finite number of sites. In the discrete case $f(q,\omega)$ actually has a more complicated structure and it only reduces to Eq.~(\ref{ff}) in the thermodynamic limit. In order not to overload the figure we choose a somewhat lower value for $N$ than in Fig.~\ref{fig-1}. We see in Fig.~\ref{fig-4} the graphical solution of Eq.~(\ref{disper}) for $N=80$. The structure is very similar to other schematic RPA models (see \textit{e.g.}~\cite{RS}) with the only difference that the smooth effective interaction is momentum and energy dependent what stems from the fact that here we have a 3-rank separable interaction instead of the usual one rank separability. The collective state gets detached from the more or less unshifted ph-roots precisely because it is situated at the edge of the continuum. One also can make a guess what may happen in the exact case. Due to screening the corresponding effective interaction $f(q,\omega)$  will bend much more strongly at the lower edge of the ph-continuum in such a way that the collective state sticks closely to the last lower vertical asymptote. 

The similarity of the RPA structure function with the exact one can also be inferred from studying the corresponding energy weighted sum rules. Let us define 
\begin{subeqnarray}
\frac{1}{2\pi}\int^{\infty}_{0}\omega S_{zz}(q,\omega)d\omega &=& -\frac{2 E_{gs}}{3 N}(1-\cos(q)) \\\slabel{rsa}
\frac{1}{2\pi}\int^{\infty}_{0}\omega S^{(2)}_{zz}(q,\omega)d\omega &=&\frac{C}{\pi}\kappa_{0}(1-\cos(q)) \\\slabel{rsb}
\frac{1}{2\pi}\int^{\infty}_{0}\omega S^{RPA}_{zz}(q,\omega)d\omega &=&\frac{1}{\pi}(1-\cos(q))\slabel{rsc}
\end{subeqnarray}
with $C=0.72221...$, $\kappa_{0}=0.9163...$ and $E_{gs}=-N(\log2 -\frac{1}{4})$ being the exact 
ground-state energy. $S_{zz}(q,\omega)$ is the exact dynamic spin structure 
factor, $S^{(2)}_{zz}(q,\omega)$ is the exact 2-spinon part of the exact dynamic spin structure 
factor~\cite{boug}, and $S^{RPA}(q,\omega)$ is the corresponding structure function in the present RPA approach. 
For $\displaystyle q=\frac{\pi}{2}$, in RPA, the sum rule gives 0.318309 and the contribution 
to the sum rule from the discrete state is 0.226843 ($71 \%$) and from the continuum it is 0.091467 ($29 \%$).
For $\displaystyle q=\frac{3\pi}{4}$, in RPA, the sum rule gives 0.543389 and the contribution 
to the sum rule from the discrete state is 0.244919 ($45 \%$) and from the continuum it is 0.298470 ($55 \%$). 
From the above we see that the RPA energy weighted sum rule Eq.~(\ref{rsc}) within $\sim 50 \%$ of the exact 
two-spinon result Eq.~(\ref{rsb}) and within $\sim 8 \%$ of the exact result Eq.~(\ref{rsa}). This latter result again underlines the semi-quantitative correct behavior of the RPA.

On Figs.~\ref{fig-1} and~\ref{fig-2} we also see that RPA actually becomes unstable for 
$q>q_{c}\simeq 0.91\pi$ because the discrete branch touches zero at $q\simeq q_{c}$. 
An RPA eigenvalue coming down to zero inevitably means that the corresponding HF 
stability matrix~\cite{RS, tho} 
\beq
{\mathcal S}(q)=\pmatrix{A & B \cr
                         B & A}
\eeq
also has a zero eigenvalue at the critical value $q=q_{c}$ and negative 
eigenvalues for $q > q_{c}$, clearly indicating that the translationally 
invariant JW-HF solution is unstable. 


The reason why the RPA performs relatively well is probably 
precisely because we forced the system to remain in the symmetry unbroken, \textit{i.e.} 'spherical', 
phase as it is well known that in 1D no spontaneously symmetry broken phase exists. 
None the less, we were curious to see what HF and HF-RPA gives in the symmetry broken phase. 
Not unexpectedly we will find that deformed HF still lowers the energy (see table~\ref{tab1}) but HF-RPA leads to an 
unphysical excitation spectrum. Follow some details of our procedure. 


We seek a new HF basis with a stable minimum in introducing new quasiparticle
operators which are a superposition of particle (p) and hole (h) operators in the 
old basis, that is~\cite{auer}
\bea
\alpha^{\dagger}_{h+\pi} &=& u_{h} \psi^{\dagger}_{h+\pi} +v_{h} \psi^{\dagger}_{h} \cr
\alpha_{h} &=& u_{h} \psi^{\dagger}_{h} -v_{h} \psi^{\dagger}_{h+\pi}
\label{iso1}
\eea
with $ u^{2}_{h}+v^{2}_{h}=1 $. Our ansatz~Eq.(\ref{iso1}) is motivated by the fact that 
the mode which becomes unstable occurs at $q=\pi$ and therefore, suggests the order parameter 
$\langle\psi^{\dagger}_{h+\pi}\psi_{h}\rangle \neq 0$, involving breaking of translational invariance.
Following the standard procedure for obtaining the amplitudes $u_{h}$, $v_{h}$ in Eq.(\ref{iso1}) 
we express the HF-energy in the new deformed state
\bea
\tilde{E}^{HF}_{0}&=&\frac{gN}{4}+\sum_{h}\left(\epsilon^{o}_{h}+\frac{1}{2}\sum_{h'}\bar{v}_{h,h',h,h'}\right) \cr
&+&\sum_{h}\bigg(\epsilon^{o}_{h+\pi}-\epsilon^{o}_{h}+\sum_{h'}\bigg\{\bar{v}_{h+\pi,h',h+\pi,h'}-\bar{v}_{h,h',h,h'} \cr
&+&\frac{1}{2}\left(\bar{v}_{h+\pi,h'+\pi,h+\pi,h'+\pi}+\bar{v}_{h,h',h,h'}
-2 \bar{v}_{h+\pi,h',h+\pi,h'}\right)v^{2}_{h'}\bigg\}v^{2}_{h}\bigg)\cr
&+&\sum_{hh'}\left(\bar{v}_{h+\pi,h'+\pi,h,h'}+\bar{v}_{h+\pi,h',h,h'+\pi}\right)u_{h}v_{h}u_{h'}v_{h'} 
\eea
and minimize with respect to $u_{h}$, $v_{h}$. We obtain
\beq
\begin{array}{c}
u^{2}_{h} \\ 
v^{2}_{h}\end{array}\bigg\}
=\frac{1}{2}\left(1 \pm \frac{\xi_{h}}{\sqrt{\xi^{2}_{h}+\Delta^{2}_{h}}}\right)
\label{par}
\eeq
where,
\bea
\xi_{h}&=&\epsilon^{o}_{h+\pi}-\epsilon^{o}_{h}
+\sum_{h'\neq h}\bigg\{\bigg(\bar{v}_{h+\pi,h'+\pi,h+\pi,h'+\pi}+\bar{v}_{h,h',h,h'} \cr
&-&\bar{v}_{h+\pi,h',h+\pi,h'}-\bar{v}_{h,h'+\pi,h,h'+\pi}\bigg)v^{2}_{h'} \cr
&+&\bar{v}_{h+\pi,h',h+\pi,h'}-\bar{v}_{h,h',h,h'}\bigg\} \cr
\Delta_{h} &=&-2\sum_{h'\neq h}\left(\bar{v}_{h+\pi,h'+\pi,h,h'}
-\bar{v}_{h+\pi,h',h'+\pi,h}\right)u_{h'}v_{h'}
\label{del}
\eea
With Eq.(\ref{par}) and Eq.(\ref{del}) we get the self-consistent 'gap' equation
\bea
\Delta_{h} &=&-2\sum_{h'\neq h}\left(\bar{v}_{h+\pi,h'+\pi,h,h'}-\bar{v}_{h+\pi,h',h'+\pi,h}\right)
\frac{\Delta_{h'}}{2\sqrt{\xi^{2}_{h'}+\Delta^{2}_{h'}}} \cr
&=&\frac{4g}{N} \sum_{h'\neq h}\frac{\Delta_{h'}}{\sqrt{\xi^{2}_{h'}+\Delta^{2}_{h'}}} 
\label{gap}
\eea
Solving the gap equation numerically allows us to calculate the new 
HF energy as a function of $g$. For $N \rightarrow \infty$, Eq.~(\ref{gap})
gives a nontrivial solution even for $g\rightarrow 0$.

\begin{figure}[htb]
\centering
\vspace{0.5 in}
\includegraphics[width=0.75\linewidth]{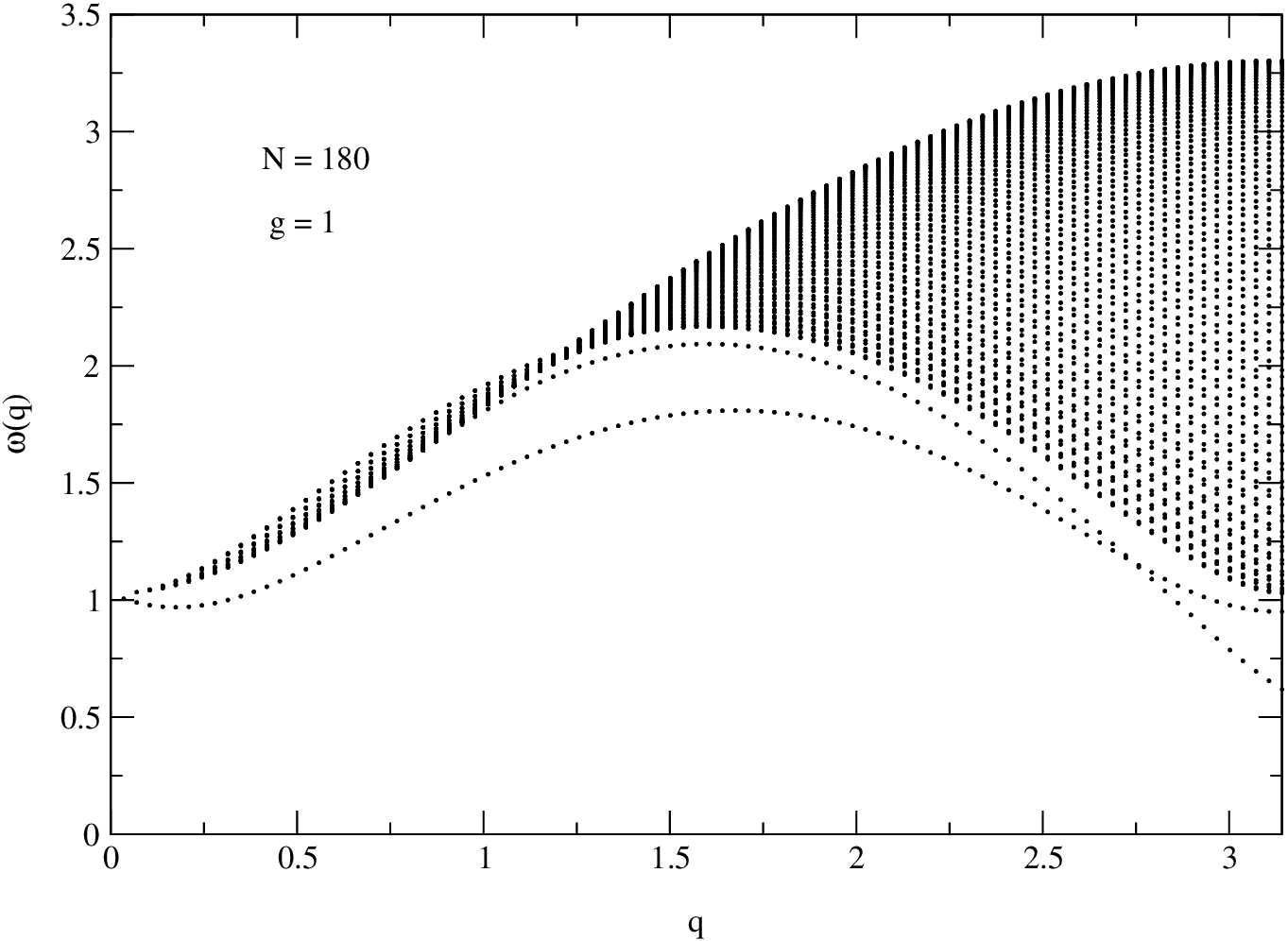}
\caption{RPA excitation spectrum in deformed basis in the case of $N=180$.}
\label{fig-5}
\end{figure}

As a last step we calculate the RPA in the new basis. For this we introduce 
in analogy to Eq.~(\ref{q1}) an RPA excitation operator of the following form
\beq
Q^{\dagger}_{\nu,q}=\sum_{ph}\tilde{X}^{\nu}_{ph}\alpha^{\dagger}_{p}\alpha^{\dagger}_{h}
-\tilde{Y}^{\nu}_{ph}\alpha_{h}\alpha_{p}, \quad q=|p-h| 
\label{}
\eeq
Proceeding in a similar way to what we have done in the spherical basis we can 
find the amplitudes $\tilde{X}, \tilde{Y}$ from corresponding RPA equations. 
We show the spectrum in Fig.~\ref{fig-5}. We see that the spectrum now contains 
a gap corresponding to Eq.~(\ref{del}) and is qualitatively similar to the exact solutions 
for an anisotropic Heisenberg Hamiltonian ($g\neq1$)~\cite{mikes}.

The fact that the spectrum in Fig.~\ref{fig-5} does not resemble much the exact one 
is not completely surprising, since we know that it is built on qualitatively wrong 
(symmetry broken) groundstate. However, the relatively good performance of the HF-RPA 
scheme in the symmetry unbroken phase motivates us to go beyond the standard RPA approach 
in applying symmetry restoration techniques~\cite{RS} and Self-Consistent RPA 
(SCRPA)~\cite{scrp2}. The latter has already produced an interesting result for the 
1D-AFH model in Ref.~\cite{kru}. 

\begin{figure}[htb]
\centering
\vspace{0.5 in}
\includegraphics[width=0.75\linewidth]{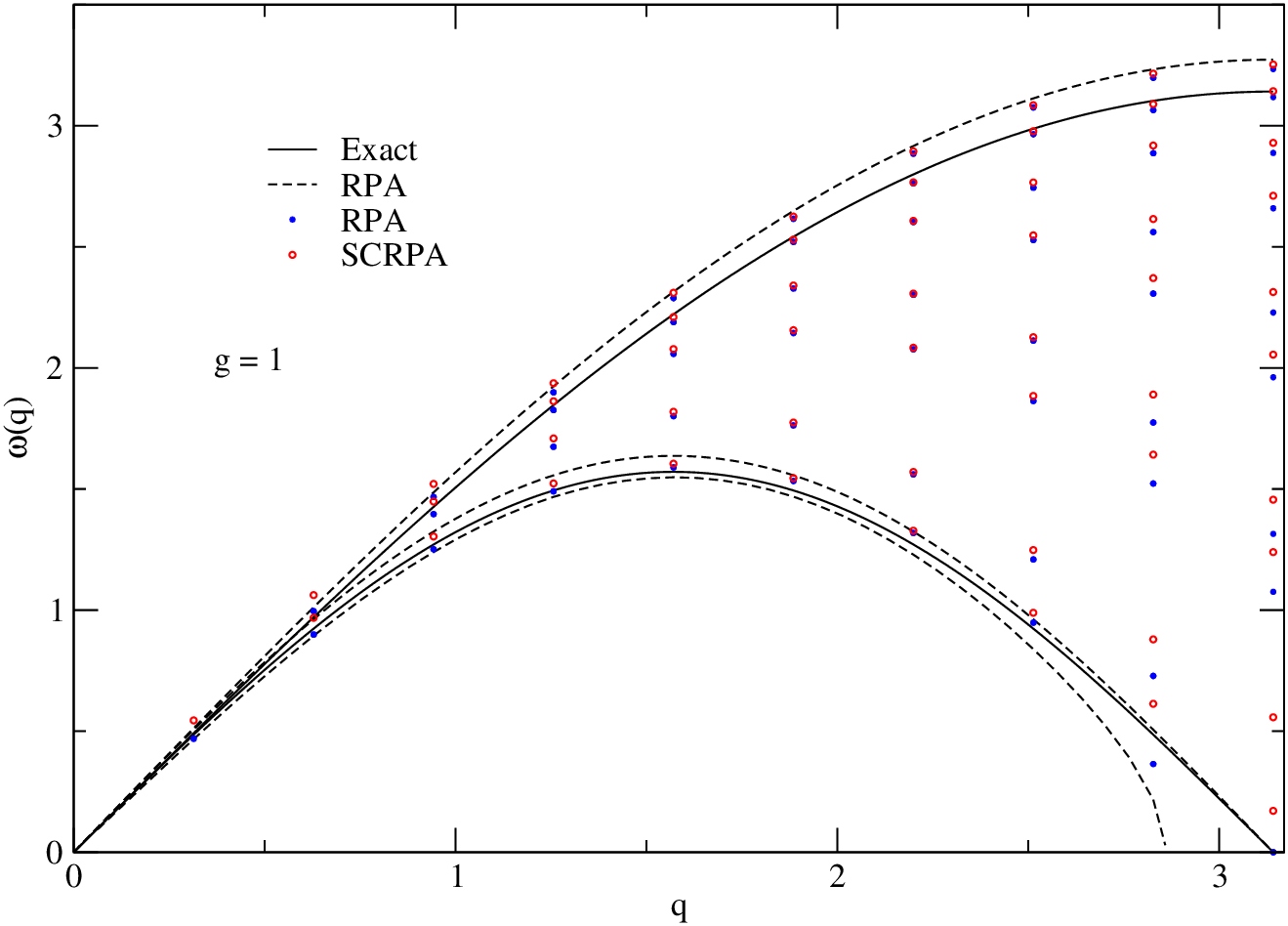}
\caption{Excitation spectrum $\omega(q)$ as function of the momentum transfer for $N=20$ (also for the thermodynamic limit). For the exact result, in the thermodynamic limit, we present the lower and upper boundary of the spectrum, they are indicated in black color (solid line). For the RPA results, in the thermodynamic limit, we present the collective mode and both boundaries of the continuum states; they are indicated in black color (dashed line). In the discrete case, RPA results are indicated by blue stars and SCRPA results by red circles.}
\label{fig-8}
\end{figure}

In this work we made a preliminary application of SCRPA to some cases of finite number of sites. Without going into details, let us shortly repeat the principles of SCRPA. It essentially consists in evaluating Eq.~(\ref{m1}) not with the HF ground state as in standard RPA but with a ground state containing RPA correlations, as this is the objective proper of RPA. In this exploratory application to the AFH model, we found it most convenient to solve the vacuum condition $Q_{\nu}|0\rangle=0$, mentioned above, pertubatively. Neglecting 3p-3h and higher configurations one obtains~\cite{Delion} :

\beq
|0\rangle \sim \left(1 + \frac{1}{4}\sum_{p_{1},h_{1},p_{2},h_{2}} z_{p_{1},h_{1},p_{2},h_{2}} \psi^{\dagger}_{p_{1}}\psi_{h_{1}} \psi^{\dagger}_{p_{2}} \psi_{h_{2}}\right)|\mbox{HF}\rangle
\label{shf}
\eeq
with
\beq
z_{p_{1}h_{1},p_{2}h_{2}}= [YX^{-1}]_{p_{1}h_{1},p_{2}h_{2}}
\eeq
\begin{figure}[htb]
\centering
\vspace{0.5 in}
\includegraphics[width=0.75\textwidth]{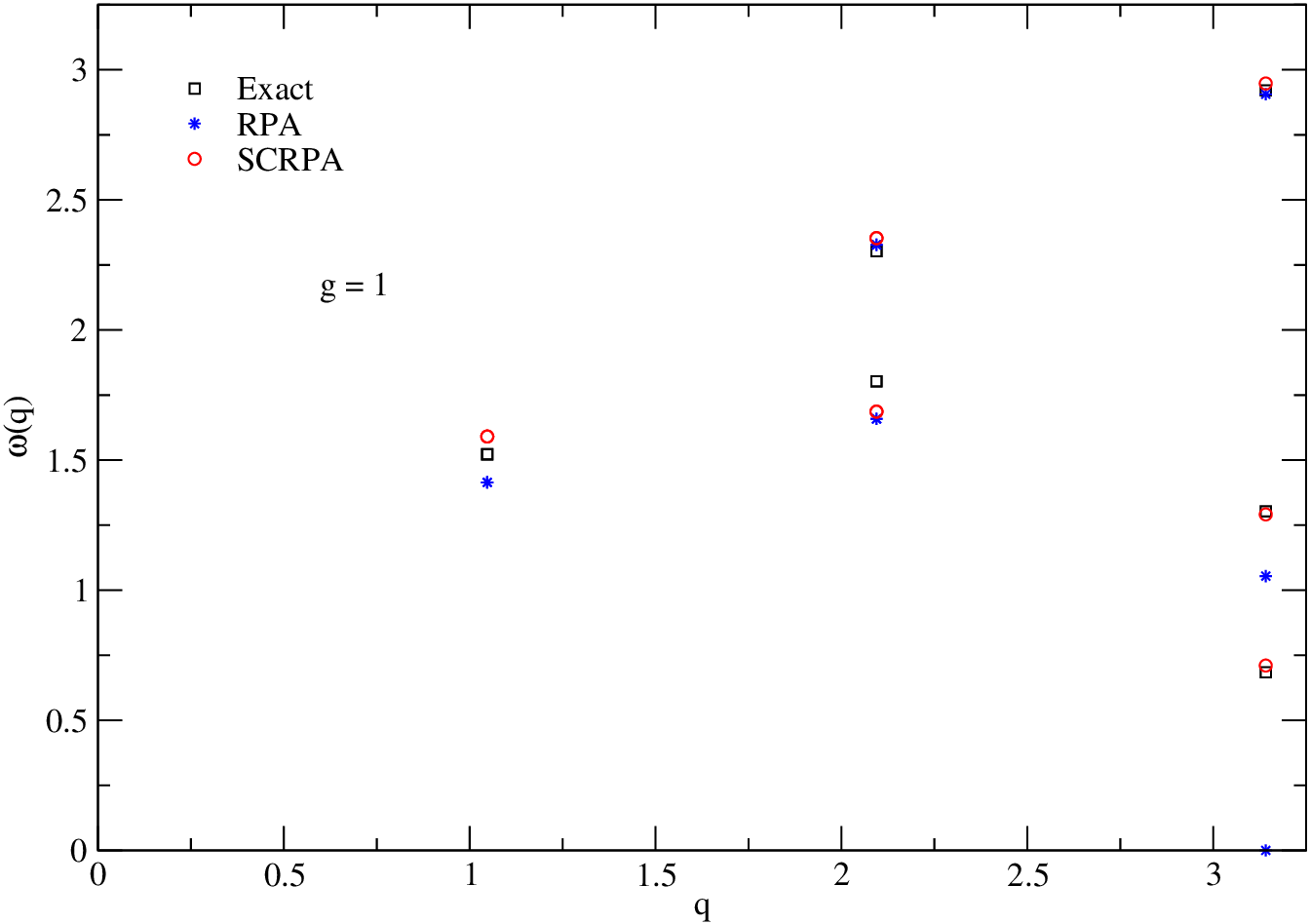}
\caption{Excitation spectrum $\omega(q)$ as function of the momentum transfer, for $N=6$. 
Exact results are indicated by black squares, RPA results by blue stars and SCRPA results by red circles.}
\label{fig-6}
\end{figure}
\begin{figure}[htb]
\centering
\vspace{0.5 in}
\includegraphics[width=0.75\textwidth]{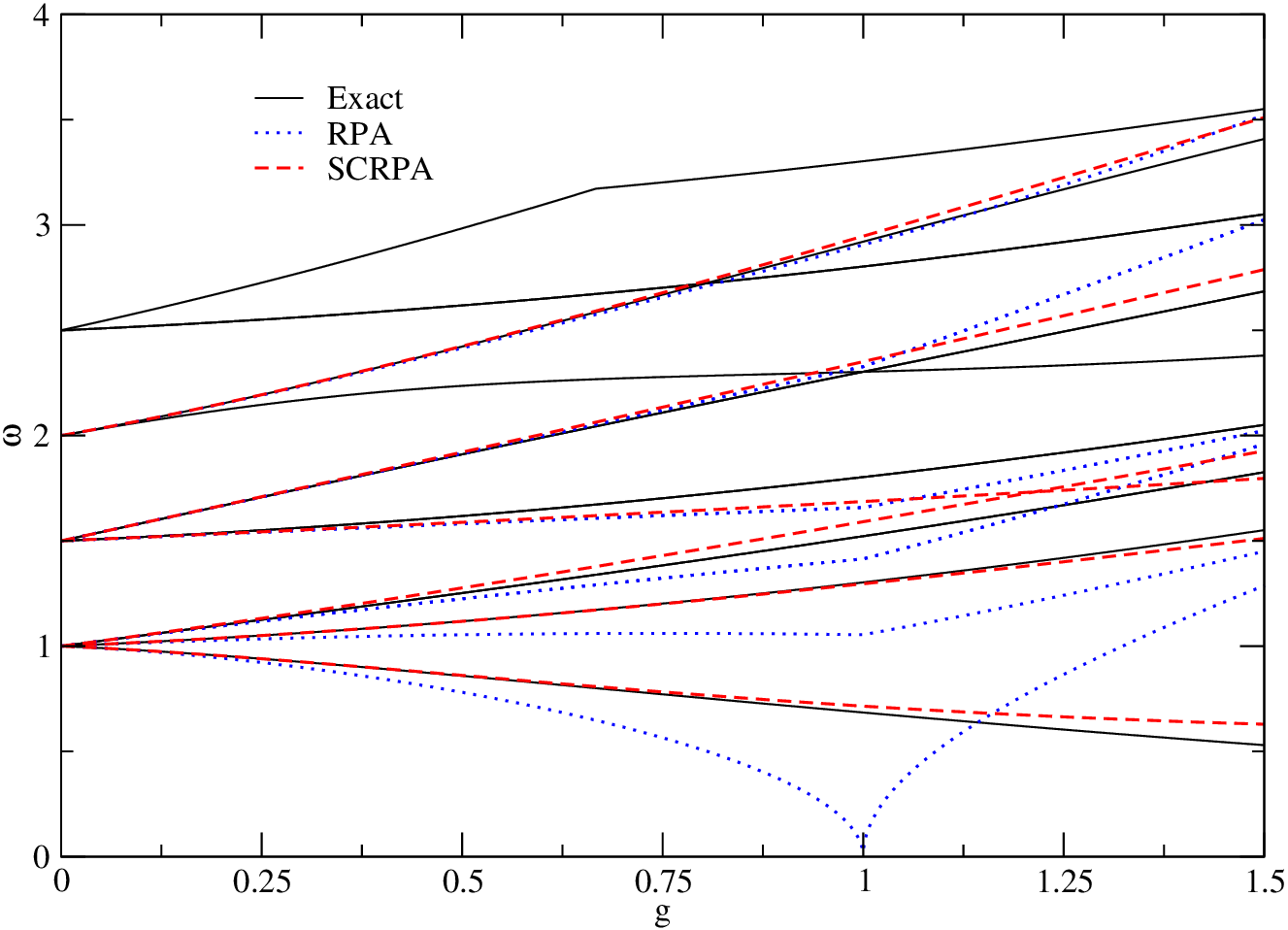}
\caption{Excitation spectrum $\omega(q)$ as function of the anisotropy $g$, for $N=6$ and for all momentum transfer. Exact results are indicated by black solid line, RPA results by blue dotted line and SCRPA results by red dashed line. For $g>1$ the RPA results are obtained in the deformed basis.}
\label{fig-7}
\end{figure}
Evaluating Eq.~(\ref{m1}) with Eq.~(\ref{shf}) yields an RPA matrix which depends in a non-linear way on the amplitudes $X$, $Y$. Solving this non-linear system of equations is numerically not completely trivial and consequently in a first attempt we restricted ourselves to quite small system sizes. In Fig~\ref{fig-6} we compare
the results from an exact diagonalization for the six sites problem with half filling with the those from SCRPA. We see that there is close agreement and a strong improvement of SCRPA over standard RPA. In Fig.~\ref{fig-7} we show the same but as function of the coupling constant g. Again we see the very important improvement of the SCRPA specially around the phase transition point $g=1$. In Fig~\ref{fig-8} we also show the SCRPA results for the case of 20-sites at half filling. We see that the lowest branch in SCRPA is strongly lifted up with respect to standard RPA. Though we do not have the exact solution at hand for this case, our results may indicate that in SCRPA no artificial gap is present any longer in the thermodynamic limit. In order to confirm this we have to solve SCRPA in the thermodynamic limit or, at least, approaching it. This shall be an investigation for the near future.

In summary, we evaluated for the first time the full RPA solution for the 1D Anti-Ferromagnetic Heisenberg model in the Jordan-Winger representation. It is shown that one obtains interesting results for spectral functions and sum rules in the symmetry unbroken phase in spite of the fact that a low-lying discrete state gets (slightly) detached from the continuum what is unphysical. The not translationally invariant Hartree-Fock solution still lowers the energy but the corresponding RPA shows, not unexpectedly, a strong artificial gap in the spectrum.
The encouraging results of standard RPA motivated us to make a first application of an improved version of RPA, the so-called Self-Consistent RPA (SCRPA)~\cite{scrp2} for the time being only for small systems.
Strong improvement of the results of SCRPA over standard RPA can be observed. One may speculate that SCRPA cures the artificial gap problem seen in Fig.~\ref{fig-3} for the structure function in the thermodynamic limit. The latter is, however, a non trivial numerical problem in SCRPA which we will try to solve in future work.
\\

{\large \bf Acknowledgments} \\

We thank H.-J. Mikeska, J. Villain, T. Ziman for useful discussions and comments.
One of us (A.R.) specially acknowledges many useful and elucidating discussions 
with C. Provid\^encia. He also is greatful to the members of theoretical group of 
the Institut de Physique Nucl\'eaire de Lyon for their hospitality at the beginning 
of this work. One of us (P.S.) is greatful to D. Foerster for an early collaboration on 
the present subject. This work has been supported by a grant provided by Funda\c c\~ao 
para a Ci\^encia e a Tecnologia SFRH/BPD/14831/2003.

\end{document}